\documentclass[12pt]{article}
\usepackage{graphicx}
\usepackage{fullpage}
\usepackage{natbib}
\title{Cartesian Coordinate, Oblique Boundary,\\
 Finite Differences and Interpolation}
\author{I H Hutchinson}
\date{Plasma Science and Fusion Center and\\ Department of Nuclear
  Science and Engineering,\\ Massachusetts Institute of Technology,
  Cambridge, MA, USA}

\begin{document}
\maketitle

\begin{abstract}
  A numerical scheme is described for accurately accommodating
  oblique, non-aligned, boundaries, on a three-dimensional cartesian
  grid.  The scheme gives second-order accuracy in the solution for
  potential of Poisson's equation using compact difference stencils
  involving only nearest neighbors. Implementation for general
  ``Robin'' boundary conditions and for boundaries between media of
  different dielectric constant for arbitrary-shaped regions is
  described in detail. The scheme also provides for the interpolation
  of field (potential gradient) which, despite first-order peak errors
  immediately adjacent to the boundaries, has overall second order
  accuracy, and thus provides with good accuracy what is required in
  particle-in-cell codes: the force. Numerical tests on the
  implementation confirm the scalings and the accuracy.
\end{abstract}

\section{Introduction}

For certain types of modeling problem it is advantageous to use
cartesian finite-differences on a lattice that does not conform to the
interfaces or boundaries of the space under consideration. Instead of
the common adoption of an unstructured mesh that conforms to the
geometry, the interfaces are considered to be arbitrary surface
shapes, and the difference equations are modified adjacent to them to
account for their boundary conditions. Such a choice is natural, for
example, if the interfaces actually move in time, since then the
remeshing cost might become excessive. 

One option for cartesian finite difference is to regard the interfaces
as being approximated by ``staircase'' contours that follow cell
boundaries. The problem with such a choice is that it constitutes a
gross approximation for smooth, possibly curved, diagonal interfaces. 
In view of the rapidly increasing computational cost of reducing the
cell size in multiple dimensions so as to reduce the errors in such a
representation, a more attractive option appears to
be to accommodate diagonal interfaces with a difference scheme and
appropriate interpolation that is higher order, notably quadratic, in
the cell size, but depends as far as possible only on the local values
of the potential.

This paper describes a scheme for solving Poisson's equation with a
wide range of general boundary conditions at interfaces, and
subsequently interpolating the potential and field
(potential-gradient) back to arbitrary points in the solution region.
The scheme has been built into a 3-D Particle-in-Cell (PIC) code for
solving self-consistent electrostatic plasma problems\cite{hutchinson11}. Examples to
demonstrate the convergence have been studied.

\section{One-dimensional differences and interpolation}\label{oneD}

\subsection{Difference approximations}

\begin{figure}[ht]
  \begin{center}\includegraphics[height=6cm]{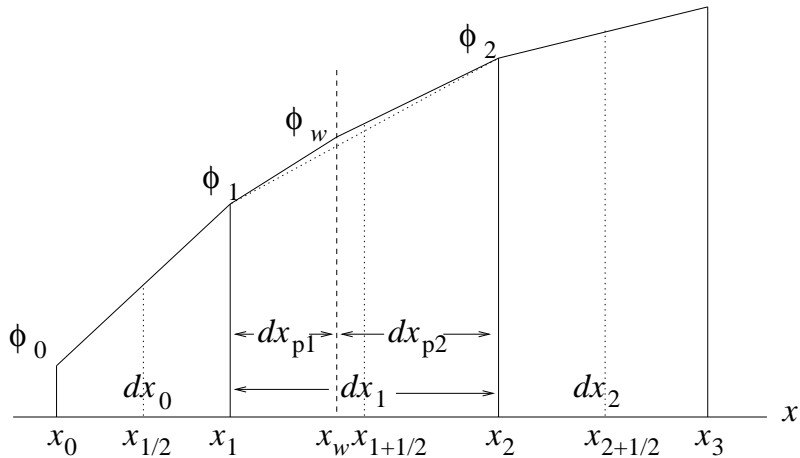}\end{center}
  \caption{Mesh notation in the vicinity of an interface cutting a
    cell at $x_w$.\label{cutcell}}
\end{figure}

First consider a one-dimensional mesh (not necessarily equally spaced)
of nodes $x_i$ on which a potential quantity $\phi$ is defined.  For
cells not affected by interfaces, we consider the gradient to be given
at positions half-way between nodes, $x_{i+1/2}\equiv
(x_i+x_{i+1})/2$, as $\phi'_{i+1/2}=(\phi_{i+1}-\phi_i)/dx_i$, where
$dx_i\equiv x_{i+1}-x_{i}$ and to be interpolated linearly between
adjacent values. See Fig.\ \ref{cutcell}. So for cells that are not
affected by interfaces, $\phi'(x)
=[\phi'_{i+1/2}(x-x_{i-1/2})+\phi'_{i-1/2}(x_{i+1/2}
  -x)]/(x_{i+1/2}-x_{i-1/2})$. The second derivative is then $\phi''_i
= (\phi'_{i+1/2}-\phi'_{i-1/2})2/(dx_i+dx_{i-1})$, which we note is
uniform because this is a second-order interpolation of
$\phi$. Because $\phi''$ is uniform, it introduces no inconsistency to
regard the $\phi''$ registration as being at position $x_i$, the same
place as $\phi_i$, even if the mesh is non-uniform so that $x_i$ is
not half way between $x_{i-1/2}$ and $x_{i+1/2}$. 

In order to alleviate notation complexities we consider specifically
cell 1 to be cut by an interface (wall) with notation illustrated in
Figure \ref{cutcell} at position $x_w$, where a boundary condition is
applied. We will consider the interpolation and solution on the left
of the interface, but the right is (under most conditions) simply the
mirror image. We are seeking a solution of the potential on the whole
grid: both sides of the interface. 

\subsubsection{Robin Boundary Condition}
The boundary condition
to be applied at $x_w$ we will take initially as being given by
\begin{equation}\label{abcbdy}
  A\phi_w+B\phi'_w+C = 0.
\end{equation}
This form, sometimes called the ``Robin'' boundary condition,
accommodates setting the value of $\phi$, or $\phi'$, or some
logarithmic gradient $\phi'/\phi$.

In bulk mesh branches (not affected by interfaces), the value of
$\phi''_i$ applies only up to $x_{i+1/2}$. If an interface intersects
the mesh between $x_i$ and $x_{i+1/2}$, as illustrated in
Fig.\ \ref{cutcell}, then $\phi''_i$ applies only up to that interface
position, $x_w$, and $\phi''_i$ must be evaluated using information
from the boundary condition, not $x_{i+1}$, because $x_{i+1}$ is not
part of the solution to the left of the interface. If an interface
intersects the mesh between $x_{i+1/2}$ and $x_{i+1}$, then again
calculation of $\phi''_i$ cannot use $x_{i+1}$, and also $\phi''$
beyond this half-cell position cannot use standard differences and
the values at $x_{i+1,2}$ because they are not part of the solution
region; so we must make some other assumption.  The most natural is to
assume that the value $\phi''_i$ applies right up to the interface,
$x_w$, no matter its distance. Consequently, the second derivative
expression,
\begin{equation}\label{phippwall}
\phi''_1 =
\left({\phi_w-\phi_1\over dx_{p1}}-{\phi_1-\phi_0\over dx_0}\right)
{2\over  dx_0+dx_{p1}}  ,
\end{equation}
must be considered to apply right up to $x_w$. And in particular
\begin{equation}\label{phipwall}
\phi'_w = {\phi_w-\phi_1\over dx_{p1}} + \phi''_1 {dx_{p1}\over 2} =  
\left[{dx_0+2dx_{p1}\over dx_{p1}}(\phi_w-\phi_1) - 
{dx_{p1}\over dx_0}(\phi_1-\phi_0)\right]{1\over dx_0+dx_{p1}}.
\end{equation}
We then determine $\phi_w$ and $\phi'_w$ from a simultaneous solution
of (\ref{abcbdy}) and (\ref{phipwall}), and substitute into
(\ref{phippwall}) to determine the second derivative, which after some
algebra can be written:
\begin{equation}\label{phippbdy}
  \phi''_1 = \left[{(-C-A\phi_1)\over Adx_{p1}+B} - {\phi_1-\phi_0\over dx_0}
\right]{2\over dx_0 + dx_{p} } ,
\end{equation}
where 
\begin{equation}\label{dxpABC}
dx_p = dx_{p1} {(Adx_{p1}+2B)\over(Adx_{p1}+B)}  
\end{equation}
can be thought of as the equivalent mesh point distance corresponding
to the boundary form.  For $B=0$ ($\phi_w$ specified) $dx_p$ is
$dx_{p1}$, while for $A=0$ ($\phi'_w$ specified) it is $2dx_{p1}$. It
can easily be verified that the corresponding expressions for
$\phi''_1$ are in agreement with intuition. This difference scheme is
entirely equivalent to that of Shortley and Weller\cite{shortley38},
and subsequently Johansen and Colella\cite{johansen98} (and others),
when used with Dirichlet conditions ($B=0$).  These schemes are
known to give $\nabla^2\phi$ to first order (truncation is $O(dx)$) for unequal
differences\cite{bramble62,jomaa10}, and hence automatically at boundaries. But because the
errors are localized at the unequal difference points, the overall
solution is generally second-order accurate\cite{matsunaga00}.

Integrating, we obtain the equivalent interpolation:
\begin{equation}\label{phiinterp}
  \phi'(x)= \phi'_{1/2} + \phi''_1(x-x_{1/2}),
\end{equation}
and 
\begin{equation}\label{phiinterp2}
  \phi(x)= \phi_1+ \phi'_{1/2}(x-x_1) + \phi''_1[(x-x_{1/2})^2-(x_1-x_{1/2})^2]/2,
\end{equation}
which can be used to approximate the potential right up to the wall. 

Notice that this scheme uses a Taylor-expansion type interpolation,
and appears to be the most accurate such scheme (for Laplace's equation
or slowly varying second-derivative) that restricts itself to a three-point
stencil. The three points are used to deduce three coefficients of the
interpolation. Potential and its derivative are continuous except
possibly at the wall.

\subsubsection{Media Boundary and Surface Charge Condition}
\label{mediasection}

A different condition is to take the interface to be the boundary
between two dielectric media with dielectric constant $\epsilon_1$ and
$\epsilon_2$. It is simple to generalize this condition to include a
specified surface charge density $\epsilon_0\sigma$ such that at the
interface, the values in the left and right regions (1 and 2) are
related by a jump condition
\begin{equation}
  \left[ \epsilon \phi'\right]_2^1 = \sigma \ .
\end{equation}
[Thus $\sigma$ is the \emph{external} charge in excess of any that
  arises from dielectric polarization.]  In order to express this
condition symmetrically, we need to extrapolate to get $\phi'_{w1}$
using (\ref{phipwall}) but we also need to extrapolate from the right
to get the gradient immediately on the other side of the interface
$\phi'_{w2}$. We do not wish to use an expression corresponding
to (\ref{phipwall}) to the right of the boundary, because that would
involve $\phi_3$, and lead to a stencil that extended more than one
leg from the node (1) under consideration. Therefore, we make the
ansatz that the charge density (i.e.\ $\epsilon \phi''$) has the same
value in region 2 as in region 1. This is consistent with our
presumption already made that $\phi''$ is uniform on one side in the
vicinity of the boundary. The resulting boundary condition is
\begin{equation}
  \epsilon_1\phi'_{w1} - \epsilon_2\phi'_{w2} = 
  \left(\epsilon_1{\phi_w-\phi_1 \over dx_{p1}} +
  \epsilon_1\phi_1'' {dx_{p1}\over2}\right) - \left(\epsilon_2{\phi_2-\phi_w \over dx_{p1}} +
  \epsilon_1\phi_1'' {dx_{p2}\over2}\right) = \sigma
\end{equation}
which is 
\begin{equation}
  (\phi_w-\phi_1)
  \left({\epsilon_1\over dx_{p1}} +{\epsilon_2\over dx_{p2}}\right)
  =
  (\phi_2-\phi_1){\epsilon_2\over dx_{p2}} - \epsilon_1\phi''_1{dx_1\over
    2} + \sigma \ .
\end{equation}
Substituting this $(\phi_w-\phi_1)$  into eq. (\ref{phippwall}) we
find
\begin{equation}\label{phippmedia}
  \phi''_1 = \left[{\sigma dx_{p2}/\epsilon_2 + (\phi_2-\phi_1) \over
      dx_{p1} + dx_{p2}\epsilon_1/\epsilon_2 } - {
    \phi_1-\phi_0 \over dx_0}\right] {2\over dx_0 + dx_p} ,
\end{equation}
where
\begin{equation}
  dx_p = dx_{p1} +  dx_1/[ 1 +
    (dx_{p1}\epsilon_2)/(dx_{p2}\epsilon_1)].
\end{equation}
This gives the difference form and the gradient extrapolation through
eqs.\ (\ref{phiinterp}, \ref{phiinterp2}). It has the merit that when
$\epsilon_1=\epsilon_2$ the differences reduce to those in the absence
of an interface, and regardless of the $\epsilon_1,\ \epsilon_2$ values
that the sum of the two $dx_p$ values adjacent to either side of the interface
is always equal to $2x_1$. Naturally $\sigma$ gives an extra term
representing a charge density.

\subsection{Difference Stencils}

In order to specify the difference scheme implementation
for Poisson's equation
\begin{equation}\label{poisson}
  \nabla^2 \phi = \rho,
\end{equation}
it is helpful to have explicit formulas for the difference stencil.
In the present approach the stencil involves only adjacent points. In
$N_d$ dimensions, this choice gives rise to a
$(2N_d+1)$-point stencil. With reference to Fig.\ \ref{cutcell}, the
difference equation at node (1) is written:
\begin{equation}
  (\sum_{\delta}P_{\delta})\phi_1 - \sum_{\delta}( Q_{\delta} \phi_{\delta}) =
  \rho_1 + \sum_{\delta} \tau_{\delta} .
\end{equation}
Here, $\delta$ takes on the values corresponding to the adjacent nodes
(negative, point 0, and positive, point 2, in the one-dimensional case
being illustrated); $P_\delta$, $Q_\delta$, and $\tau_\delta$ are
coefficients that will be specified in a moment. Note that only the
sums $\sum_{\delta}P_{\delta}$ and $\sum_{\delta} \tau_{\delta}$
need to be stored in addition to all $Q_\delta$. Thus the storage per
node required to specify the difference equation is $2N_d+2$ quantities.

Along the dimension under consideration, the value of the diagonal
coefficient is 
\begin{equation}\label{pequ}
  P = {2 \over (dx_p + dx_m)d_{eff}}.
\end{equation}
The $dx_{p,m}$-quantities can be considered to represent the distances
from the node underconsideration to the
control position in the positive $dx_p$ and negative $dx_m$
directions, and there is an extra effective distance $d_{eff}$. When there is
no boundary, for example when the negative direction is between $x_1$ and
$x_0$, then the control distance is to the adjacent node:
\begin{equation}
  dx_m = dx_0.
\end{equation}
In the direction of the boundary, the other quantities are given in
table \ref{coeftable}, now to be explained.

\begin{table}[htp]
\begin{tabular}{|l||c|c|c|c|}
\hline
Boundary  Condition & $dx_p$  & $d_{eff}$ & $Q$ & $\tau$ \\
\hline
\hline
None   & $dx_1$ & $dx_1$   & $P$ & 0 \\
\hline
{\vrule depth 8pt height 14pt width 0pt}
$A\phi+B\phi'+C=0$ & ${Adx_{p1}+2B\over Adx_{p1}+B}dx_{p1}$ &
$dx_{p1}+{B\over A}$ & 0 & $-{C\over A}P$\\
\hline
{\vrule depth 8pt height 16pt width 0pt}Continuity
 & $dx_{p1}+{dx_{p2}^2\over dx_{p1}}{ B\over Adx_{p2}+B} $ 
& $dx_{p1}$ & ${B\over Adx_{p2}+B}P$ 
&  ${-Cdx_{p2}\over Adx_{p2}+B}P$ \\
\hline
{\vrule depth 8pt height 12pt width 0pt}
Media $[\epsilon{\partial\over\partial n}\phi]_2^1 =\sigma$ &
 $dx_{p1}+dx_1/(1+{dx_{p1}\epsilon_2\over dx_{p2}\epsilon_1})$
 & $dx_{p1} + {\epsilon_1\over\epsilon_2}dx_{p2}$ &
$P$ & $-P \sigma dx_{p2}/\epsilon_2$ \\
\hline
\end{tabular}
\caption{Coefficients of the difference scheme.\label{coeftable}}
\end{table}

The ``None'' condition means a situation where there is no
boundary. This row gives the standard coefficients of the difference
scheme away from boundaries. The table specifies only the coefficients
for the positive $\delta$ (point 2 relative to point 1). The
coefficients for the negative $\delta$ (point 0 relative to point 1)
follow from reflectional symmetry. For example for the ``None'' case
the negative side coefficients are given by eq.\ (\ref{pequ}) with
$dx_m=dx_1$, $dx_p=dx_0$, $d_{eff}=dx_0$.

The Robin case, $A\phi+B\phi'+C=0$, prescribes the condition that
applies on the left side of the interface position (i.e.\ it is what
point 1 sees in the positive direction towards point 2). In this case,
the value of potential at $x_2$ is unused, and instead a constant
$\tau$ contribution is added. It acts like an additional charge
density. Together the coefficients of this row are equivalent to equation
(\ref{phippbdy}). When there is an interface on the opposite side of
the stencil center-point from the coefficients being evaluated, the value of
$dx_m$ to be used is equal to $dx_p$ for that opposite direction. In
other words, the coefficients for negative $\delta$, in the stencil centered
at $x_1$, are obtained by taking $dx_p=d_{eff}=dx_0$, and $Q=P$, but
$dx_m= {Adx_{p1}+2B\over Adx_{p1}+B}dx_{p1}$.

When a gradient boundary condition is applied using  the Robin form
with $B\not=0$, then the boundary condition applied on the other side
of the interface is usually just continuity of potential. If any other
condition is applied, including the identical Robin condition, then a
discontinuity in the potential will arise unless the exterior boundary
conditions are consistently chosen. A discontinuity in potential
\emph{gradient} is usually physically acceptable (as a surface charge) but
a potential discontinuity is usually not. Therefore a ``Continuity''
boundary condition is provided explicitly in the table to obtain the
coefficients for point (2) relative to point (1) for the case
where the Robin condition applies on the other side of the
interface. That is, when point (2) sees the Robin condition in the direction
of point (1) with the prescribed $A$, $B$, and $C$ values. Naturally,
when $B=0$, this continuity condition reverts to the fixed-$\phi$
boundary condition: it becomes identical to the row above it. 

The ``Media'' row documents the stencils needed to implement the
treatment explained in section
\ref{mediasection}.

\section{Multiple Dimensions}

The one-dimensional problem just discussed is relatively
straightforward. Extension to multiple dimensions requires that a
number of important complicating factors be considered.

\subsection{Difference Stencil and Boundary Conditions}

The difference scheme considered here for the Laplacian in $N_d$
dimensions is based upon the simplest stencil employing only the $2N_d$
adjacent points, giving a (2$N_d$+1)-point stencil. If we consider a
second-order Taylor expansion about a point that for notational
convenience we take to be the origin
of coordinates, it may be written:
\begin{equation}
  \phi(x_\alpha) = \phi_1 + \sum_\alpha^{N_d} p_\alpha x_\alpha +
  \sum_\alpha^{N_d} q_\alpha x_\alpha^2 + \sum_\alpha^{N_d}\sum_{\beta>\alpha} r_{\alpha\beta}
  x_\alpha x_\beta , 
\end{equation}
where $\alpha$ refers to the dimension number and $p$, $q$, and $r$
are the expansion's coefficients numbering $N_d$, $N_d$, and $N_d(N_d-1)/2$
respectively. The $2N_d$ differences of the (2$N_d$+1)-point stencil
permit the determination of only the $p_\alpha$ and $q_\alpha$ in addition to
$\phi_1$. The $r$ coefficients, which exist for $N_d>1$ in increasing
numbers, cannot be determined. Fortunately, those cross terms do not
contribute to the Laplacian, so that the multidimensional stencil can
still determine the Laplacian to the same truncation order, using the
1-D difference scheme for each dimension.

If the boundary conditions are of the Dirichlet type,
specifying $\phi$, then the generalization essentially
straightforward. We use the $B=0$ version of eq (\ref{phippbdy}) for
second derivatives, and stencil contributions, in all coordinate
directions. Their sum comprises the Laplacian. Such an approach in
two-dimensions is frequently called the ``Shortley-Weller''
approximation, and is known to be second-order accurate, for
essentially the same reasons as for 1-D.

\paragraph{Gradient Boundary Conditions.}

\begin{figure}[ht]
  \begin{center}
  \includegraphics[height=6cm]{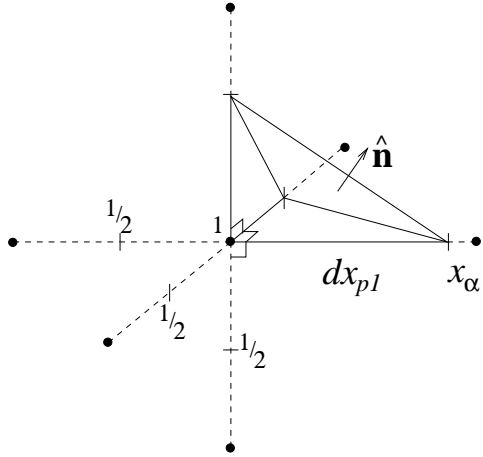}
  \caption{Interface intersecting the lattice legs forming a
    tetrahedron in 3 dimensions.\label{tetrahedron}}
  \end{center}
\end{figure}

If, however, we have a full boundary condition at the wall
of the form
\begin{equation}\label{mdbc}
  A \phi + B \hat{\bf n}. \nabla \phi + C = 0,
\end{equation}
where $\hat{\bf n}$ is the normal unit-vector, then we need to
translate this requirement into an equivalent for each of the
dimensions, where wall intersections occur between nodes. It is the
presence of the normal gradient $\partial\phi/\partial n$ that is
problematic, so Neumann conditions ($A=0$) experience the same
difficulty.  Suppose the unit vector for (cartesian) dimension
$\alpha$ is $\hat{\bf e}_\alpha$, and consider an elementary volume
(half of a hypercuboid: in 3-D a tetrahedron, see
Fig.\ \ref{tetrahedron}) defined by the planes normal to the dimension
axes cut by the wall plane with normal $\hat{\bf n}$.
 If the area of
the face of this volume defined by the wall plane is $\Omega$, then
the area of the face perpendicular to axis-$\alpha$ is $\Omega
\hat{\bf e}_\alpha.\hat{\bf n}$. Consequently flux conservation
demands that $\sum_\alpha \hat{\bf e}_\alpha.\hat{\bf n}
\nabla_\alpha\phi = \hat{\bf n}.\nabla\phi$. Therefore the
flux-conservation implicit in the Laplacian is maintained together
with eq (\ref{mdbc}) if in the $\alpha$-dimension boundary condition,
\begin{equation}
  A\phi+ B_\alpha  \nabla_\alpha\phi + C =0,
\end{equation}
we take $B_\alpha=(B/\hat{\bf
  e}_\alpha.\hat{\bf n})$, (since $\sum_\alpha (\hat{\bf
  e}_\alpha.\hat{\bf n})^2=1$). This choice amounts to taking $\nabla\phi =
\hat{\bf n} |\nabla\phi|$, and ignores contributions to
$\nabla_\alpha\phi$ arising from any \emph{tangential} component of
the gradient. Such contributions might change the values of
$\nabla_\alpha\phi$ for individual directions $\alpha$, but they sum to zero
in the normal component. In so far as the control volume adjacent to
the boundary can indeed be considered to be half a hypercuboid, the
tangential gradient \emph{does not contribute} to the divergence of
the flux, because at the interface its normal component is zero and at
the lattice cell-boundaries the flux is obtained directly from the
finite differences there, regardless of what is the flux tangential to
the oblique interface.
Thus, for multidimensional evaluation of the Laplacian, in the vicinity
of an interface, one can use for each dimension the one-dimensional
stencil derived in section \ref{oneD} but with the boundary condition
coefficient $B_\alpha=(B/\hat{\bf e}_\alpha.\hat{\bf n})$. 

By the same argument, when a surface charge density $\sigma$ is
present, which gives rise to a gradient discontinuity, the fraction of
that charge density to be attributed to dimension $\alpha$ is
${\bf\hat{e}}_\alpha.{\bf\hat{n}}$. Then, associated with an
interface area $\Omega$, dimension $\alpha$ possesses a projected area
$\Omega {\bf\hat{e}}_\alpha.{\bf\hat{n}}$. When each dimension area
is multiplied by its charge density
${\bf\hat{e}}_\alpha.{\bf\hat{n}}\sigma$ and the resultant summed over
all dimensions, the total surface charge $\Omega\sigma$ results.
 A lattice leg along the $\alpha$ dimension
intersecting the surface must use the 1-dimensional difference scheme
of eq.\ (\ref{phippmedia}), but with charge density
${\bf\hat{e}}_\alpha.{\bf\hat{n}}\sigma$.

These choices are the appropriate ones for a difference scheme with a
(2$N_d$+1)-point stencil in $N_d$-dimensions.  With the standard expression
for the second derivative along each dimension or its generalizations,
eqs.\ (\ref{phippbdy}) or (\ref{phippmedia}) in hand, (as summarized
in Table 1) one can then solve the resulting matrix equation by
standard methods such as SOR or conjugate gradient minimization, to
find the potential on the mesh.

\subsection{Multidimensional Interpolation}

More complex mathematical considerations and by far the more difficult
coding challenges lie in the back-interpolation of the solution to
arbitrary position, rather than the solution of the difference
equations.

\paragraph{Gradient interpolation.}

For the dynamics of particles moving in the field, the gradient of the
potential is the most important.  \emph{Gradient interpolation} is
done most naturally using the information embedded in the difference
stencils. For gradients in direction $\alpha$ at positions along
$\alpha$-lattice-legs, i.e.\ at coordinate values in the other
dimensions corresponding to a node, the 1-D interpolation is entirely
natural and straightforward.  Eq.\ (\ref{phiinterp}) with
eq.\ (\ref{phippbdy}) at bulk nodes and Dirichlet boundaries, or with
eq.\ (\ref{phippmedia}) at Media and surface-charge interfaces can be
used.

However, at interfaces where the Robin (with $B\ne 0$) or any
normal-gradient condition is applied, interpolation that sets the
tangential gradient to zero is insufficient. It is sufficient for
evaluating the Laplacian to ignore tangential gradient but not for the
field interpolation. We must estimate the tangential gradients ($N_d-1$
quantities). We wish to do this for convenience and compactness
\emph{without} extending our consideration beyond the standard
$(2N_d+1)$ stencil node values. We therefore have to drop cross-gradients from
the Taylor expansion as we do in the bulk scheme. That means we can
ignore the normal derivative of the tangential gradients, and
determine the gradients on the opposite side of the central node from the
interface intersections.  That is (see Fig.\ \ref{tetrahedron}) from
the gradients at positions $x_{\alpha,1/2}$, taking the tangential
field to be simply constant for the cell under
consideration. The estimate of the tangential gradient is therefore
\begin{equation}
  \nabla_t\phi = \nabla\phi|_{1/2} - (\hat{\bf n}.\nabla\phi|_{1/2}).\hat{\bf n}
\end{equation}
(where an appropriate interpolation for $\hat{\bf n}$ is used if
curvature is present). Then when the interface boundary condition is
the specification of ${\partial \phi\over\partial n}$, the gradient at
each of the face points ($dx_{p1}$) is 
\begin{equation}
  \nabla\phi|_w = \nabla_t\phi + {\partial\phi\over\partial n}
  \hat{\bf n} = \nabla\phi|_{1/2} - \left(\hat{\bf
    n}.\nabla\phi|_{1/2}\right).\hat{\bf n}\ +
{\partial\phi\over\partial n}\hat{\bf n}\ .
\end{equation}
For direction $\alpha$, therefore, 
\begin{equation}
  \phi'_{\alpha w} = \left.{\partial\phi\over\partial x_\alpha}\right|_w={\phi_1-\phi_{\alpha0}\over
    dx_{\alpha0}} - n_\alpha \left(  \sum_\beta^{N_d} n_\beta
       {\phi_1-\phi_{\beta0}\over dx_{\beta0}} -{\partial\phi\over\partial n}
\right)\ .
\end{equation}
Substituting for ${\partial\phi\over\partial n}$ from the Robin
condition we get 
\begin{equation}
  \phi'_{\alpha w} =-n_\alpha (A/B) \phi_w
 - n_\alpha \left(C/B + \sum_\beta^{N_d} n_\beta
       {\phi_1-\phi_{\beta0}\over dx_{\beta0}} -{\phi_1-\phi_{\alpha0}\over
    n_\alpha dx_{\alpha0}} 
\right)\ .
\end{equation}
This is now in the form of a Robin condition in the single ($\alpha$)
dimension, but with new Robin coefficients  
\begin{equation}
  A'=A\ ,\quad B'=B/n_\alpha\ ,\quad C'=C + B\left(\sum_\beta^{N_d} n_\beta
       {\phi_1-\phi_{\beta0}\over dx_{\beta0}} -{\phi_1-\phi_{\alpha0}\over
    n_\alpha dx_{\alpha0}} 
\right)\ .
\end{equation}
These primed coefficients can be substituted into our prior expression
for the second derivative, eq (\ref{phippbdy}), to obtain:
\begin{equation}
  \phi''_{\alpha1} =-{\left[C + A\phi_1 +
    A(\phi_1-\phi_{\alpha0}){dx_{\alpha p1}\over dx_{\alpha0}}
    + B\sum_\beta^{N_d} n_\beta
       {\phi_1-\phi_{\beta0}\over dx_{\beta0}}\right]
  \over Adx_{\alpha p 1} +B/n_\alpha} {2 \over dx_{\alpha0} + dx_p}\ ,
\end{equation}
where $dx_p = dx_{\alpha p1}(An_\alpha dx_{\alpha p1} +2B)/(An_\alpha
dx_{\alpha p1} + B)$. It is helpful to note that in the limits $B\to
0$, or $n_\alpha\to 1$ this expression is identical to
(\ref{phippbdy}).  For other parameters, the most important extra
distinction is the inclusion of the $B$ term in the numerator, which
involves values $\phi_{\beta0}$ off the $\alpha$-lattice-line.

\begin{figure}[ht]
  \begin{center}
  \includegraphics[height=7cm]{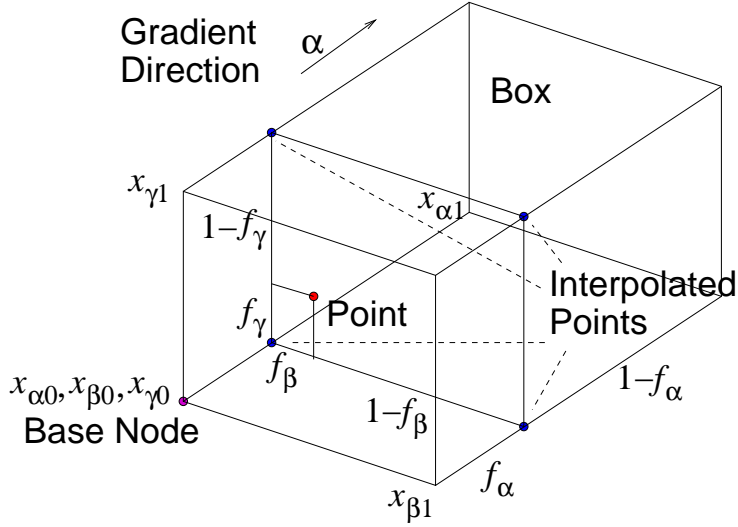}
  \caption{Gradient-interpolation geometry within an elementary
    lattice box.\label{boxinterp}}
  \end{center}
\end{figure}

\paragraph{Transverse interpolation}

Referring to Fig \ref{boxinterp} which shows an example in 3
dimensions, we use as base the node closest to the point, and the
``box'' consisting of the elementary cuboid whose vertices are
adjacent mesh nodes, in which the point lies. 

For the gradient along the $\alpha$-direction, we use the quadratic
interpolation, eq (\ref{phiinterp}), along each of the $2^{N_d-1}$ lattice
legs (i.e.\ box edges) aligned along the $\alpha$-direction, to
obtain the gradients at interpolated points where the
$\alpha$-coordinate is $x_\alpha$, a fraction $f_\alpha$ of the lattice
spacing from the base, and the other coordinates correspond to nodal
values. Then we use multilinear interpolation between those values in
the orthogonal ($N_d$-1)-dimension subspace to approximate the gradient
at the point. Multilinear interpolation can be considered to be
defined inductively via
\begin{equation}\label{multilin}
\phi'(x_\alpha,x_\beta,...,x_\eta,...x_{N_d})=
f_\eta   \phi'(x_\alpha,x_\beta,...,x_{\eta0},...x_{N0}) +
(1-f_\eta)   \phi'(x_\alpha,x_\beta,...,x_{\eta1},...x_{N1})
\end{equation}
for all $\eta \ne \alpha$. It has the important merit of giving
continuity at cell boundaries but requiring only localized box information.
This scheme provides gradient (field) estimates that are continuous
and have at least first order accuracy.

If at the first, quadratic, stage we find a node from which we would
do interpolation, e.g. $(x_{\alpha 0},x_{\beta 1},x_{\gamma0})$, to be
outside the point's solution region, i.e.\ on the other side of an
interface, then we examine the other end of the lattice leg under
consideration $(x_{\alpha 1},x_{\beta 1},x_{\gamma0})$. If that end
lies in the point's region, then we extrapolate from that instead. If
that second end does not lie in the point's region either, then the
lattice leg is entirely outside the interpolation region. In that
case, an ``over-extrapolation'' is sought by looking one more node in
either direction. If a node is found that is in the region, then
(over-)extrapolation from that node to the lattice position under
consideration is adopted. Such a value is considerably more uncertain
than values corresponding to points that are in the region or directly
extrapolated from the region. Over-extrapolated values are given a
weight $w$ equal to one minus the fractional position relative to the
lattice end adjacent to the region. That is, if position $x_{\alpha0}$
is in the region, and $x_{\alpha1}$, $x_{\alpha2}$ not, then the
weight at $x$ such that $x_{\alpha1}<x<x_{\alpha2}$ is $w(x)= 1 -
(x-x_{\alpha1})/(x_{\alpha2}-x_{\alpha1})$. If there is no
extrapolation available, then the weight is set to zero.

When one or more of the weights in a multilinear interpolation is not
unity, a weighted multilinear interpolation is used. It can be
expressed as a sum over all the $2^{(N_d-1)}$ dimensional corners of the box
within which the interpolation is being done, denoted by the vector
index whose components have the values $i_\eta=0\ {\rm or}\ 1$, where $1\le
\eta\le N_d-1$ denotes the dimension. Thus
\begin{equation}
 \phi'= \sum_{i_\eta=0,1} A(i_\eta) \phi'(i_\eta)\bigg/ \sum_{i_\eta=0,1} A(i_\eta)\quad {\rm
    where}\quad
A(i_\eta) = w(i_\eta) \prod_\eta (1-i_\eta+f_\eta(2i_\eta-1) ).
\end{equation}
When all $w=1$, this expression is equivalent to eq (\ref{multilin}).


For concave interfaces, it is theoretically possible for a particle to
be in a box which has no vertices in its region. If that happens, a
fall-back is needed, the natural thing is to choose as base node the
nearest node that is in its region. This problem can usually be
avoided by choice of mesh in relation to objects.

\paragraph{Potential interpolation.}

Because we do not directly differentiate the potential, using instead
the gradient interpolation just explained, satisfactory values
for the potential are obtained by multilinear interpolation, and do
not require higher order interpolation schemes. Multilinear interpolation is
unproblematic in the ``bulk'' where all the box vertices are valid
points in the relevant region of the grid, and no interface crossings
occur on its edges. The problematic issue is how to extrapolate to the
interface, when box vertices are absent from the valid region because
they are the other side of an interface. After some experimentation,
the following approach was found to work well using modest
computational resources.  First, all the vertices of the box are
examined. If any do not lie in the region, they must be filled in with
extrapolated values. The extrapolated values are obtained by
calculating (during this examination) the centroid, the mean value,
and the mean gradients of the ``valid'' vertices of the box (ones that
lie in the region). Additional steps must be taken for any dimension
in which the gradient is undetermined, which happens when all the
valid vertices have the same coordinate in that dimension. Such a
situation occurs when an entire face of the box is outside the valid
region.  The occurence is not all that unusual. In such a situation of
undetermined gradient, we instead determine the gradient in that
dimension, using the gradient interpolation code, evaluated at the
point. The values that are filled in to the invalid vertices are then
the mean cell value plus the dot product of the gradient and the
vector to the vertex from the centroid. Multilinear interpolation
can then proceed as usual on the entire box including the filled-in vertices.

\section{Example Accuracy Tests}

\begin{figure}[htp]
  \includegraphics[height=6cm]{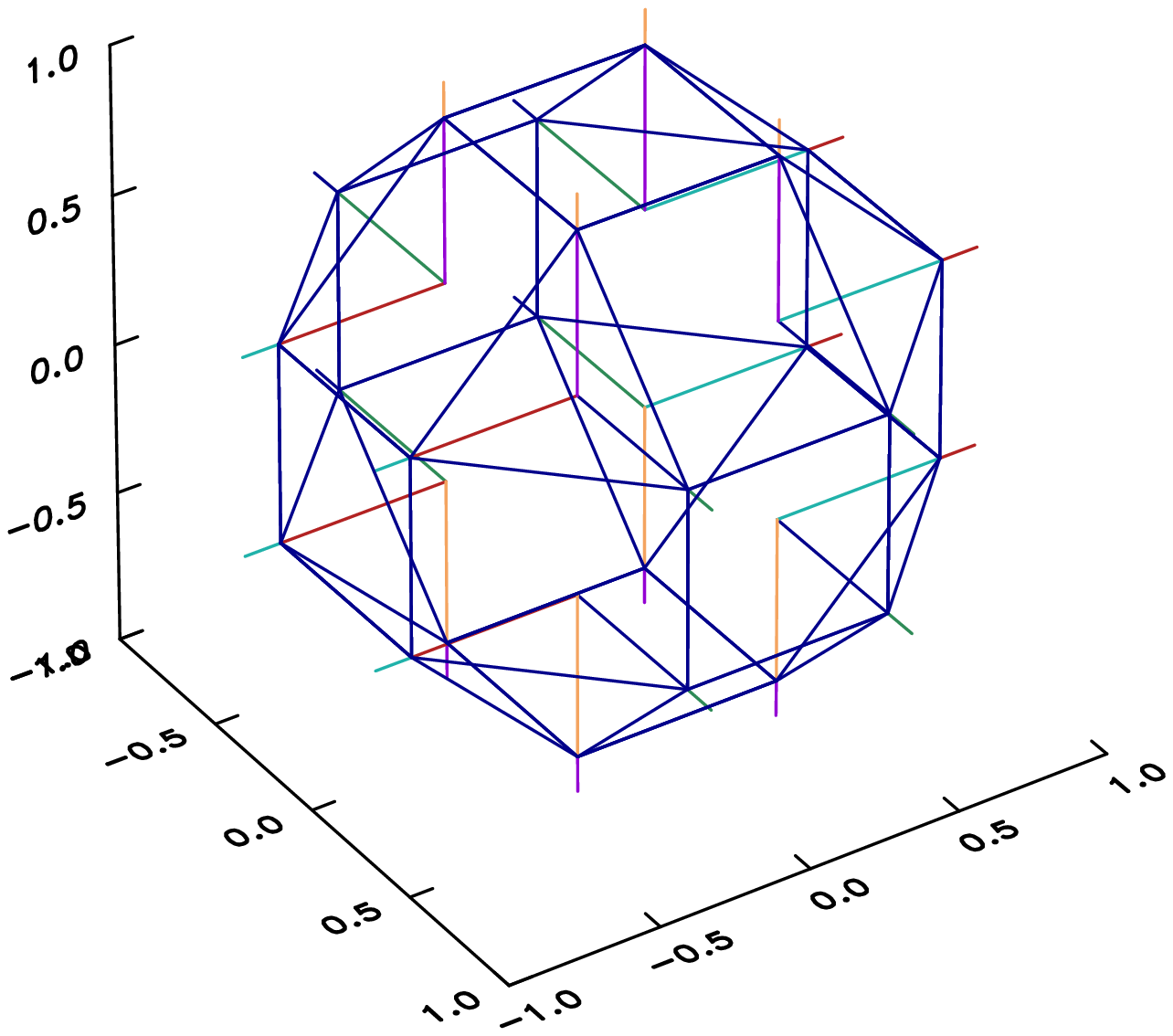}\hskip -2cm (a)\hskip 1.8cm
  \includegraphics[height=6cm]{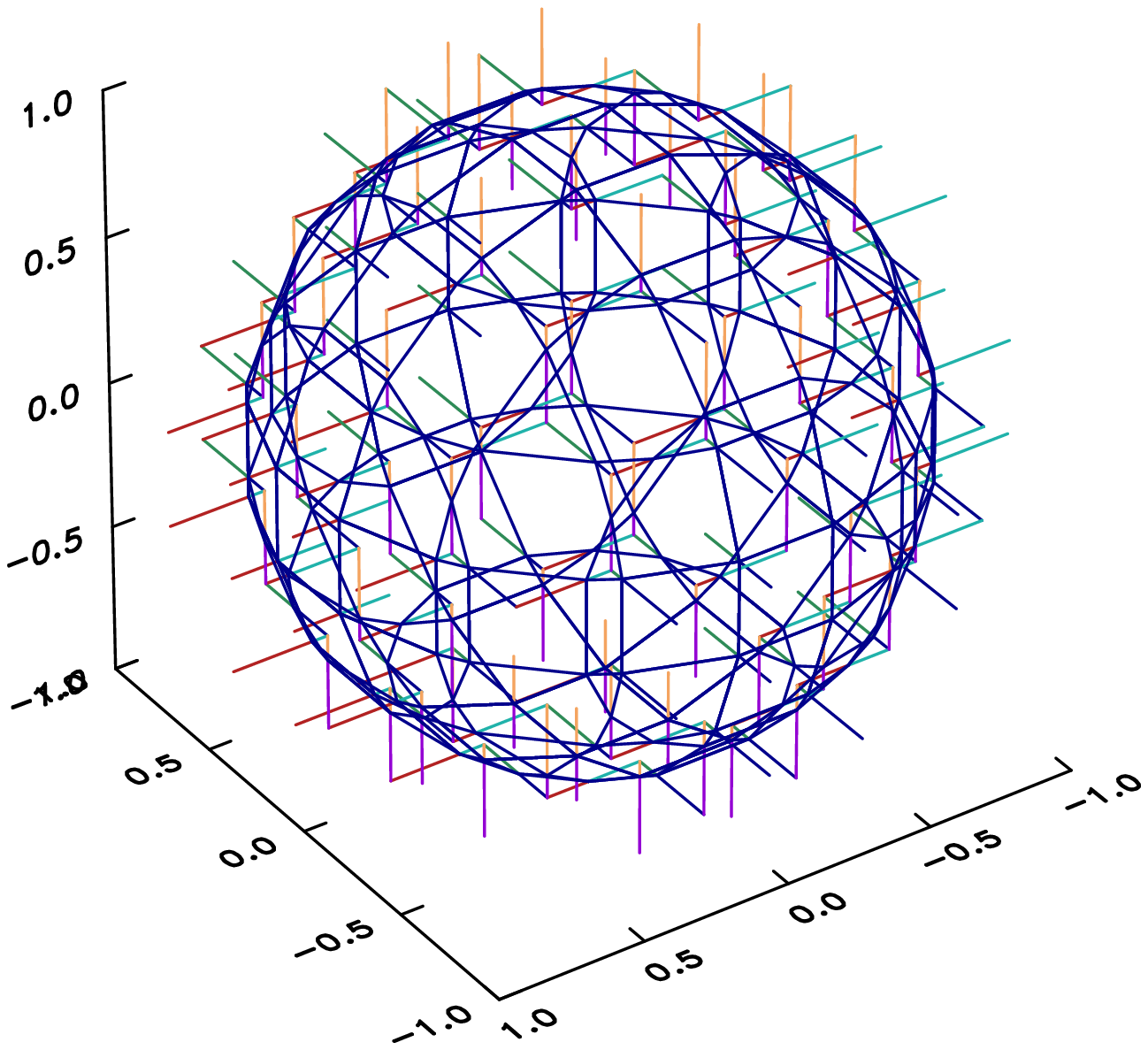}\hskip -2cm (b)
  \caption{Representation of spherical object on $16^3$ (a) and $32^3$ (b)
    grids.\label{spherestick}}
\end{figure}

Here we give examples of the typical accuracy of the difference
solution and interpolation schemes. The case illustrated is of a
sphere of fixed potential 10, radius 1, inside a second sphere of
radius 5 at which the potential logarithmic radial gradient is -1. 
It is solved on a mesh over the region $-5\le x,y,z \le 5$.
The inner sphere wireframe is shown in Fig.\ \ref{spherestick} for two
grids, $32^3$ and $16^3$.
Also shown in this figure are the lattice legs which are intercepted
by the sphere. The wireframe nodes are those intercepts. 

The solution
of this problem: $\nabla^2\phi =0$, is of course
$\phi=\phi(1)/r$. 
\begin{figure}[htbp]
  \includegraphics[width=.49\hsize]{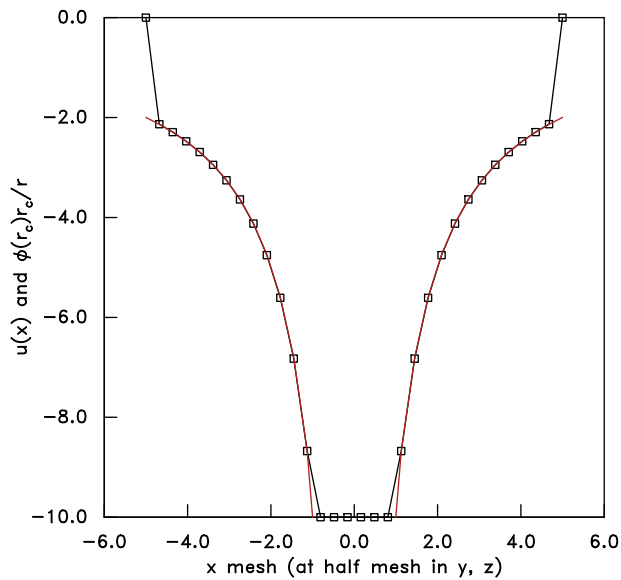}   
  \caption{Solution of the difference equation.\label{phinode}}
\end{figure}
In Fig.\ \ref{phinode} the values at the nodes in
the plane of constant $y, z$ (through the center) are
compared with this analytic solution. Inside the inner sphere the
potential is constant, dictated by continuity at the interface, as
would occur for a conducting sphere. (Outside the outer sphere the
potential is set to zero.)

\begin{figure}[ht]
  \includegraphics[width=.49\hsize]{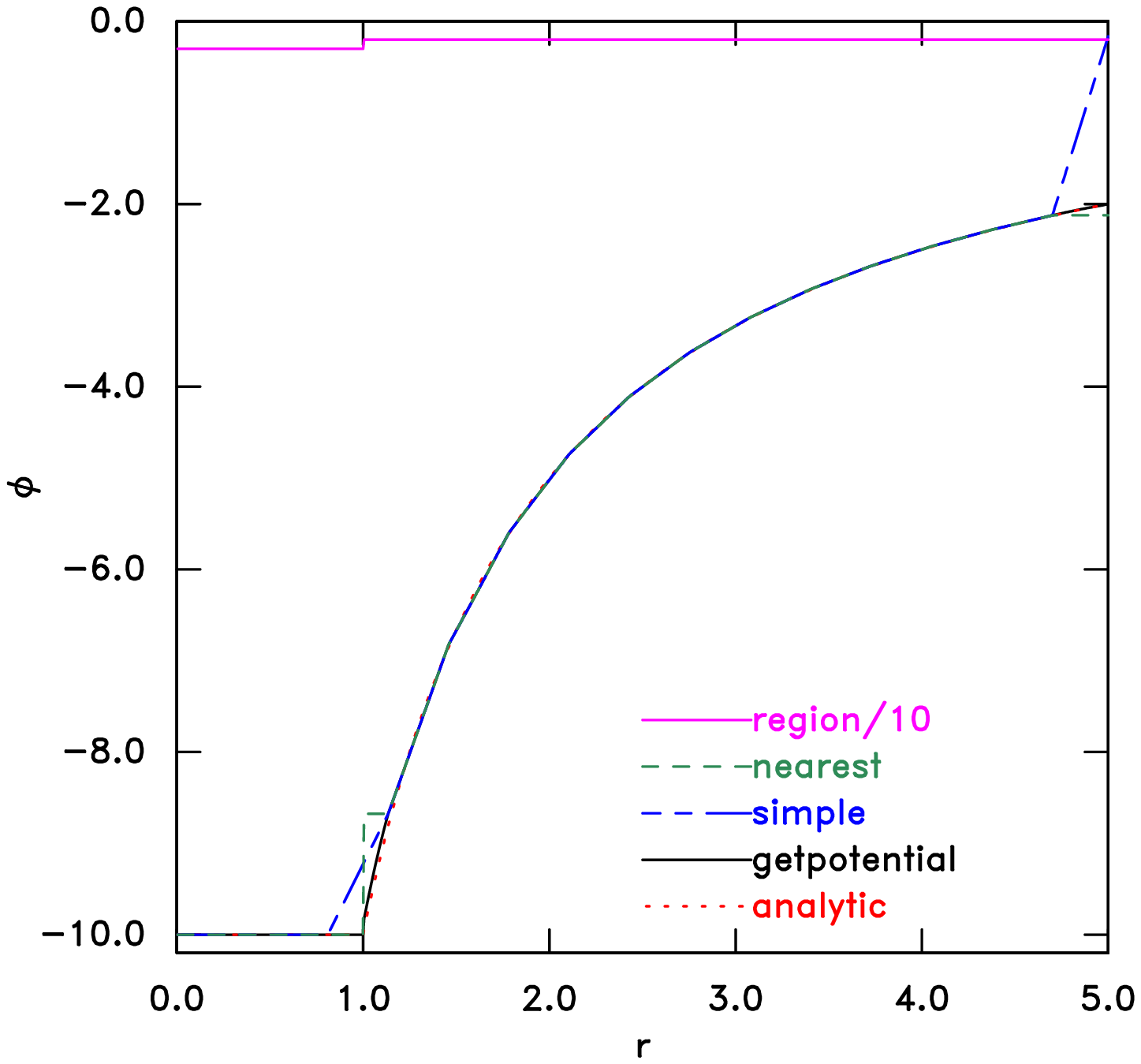}   
  \includegraphics[width=.49\hsize]{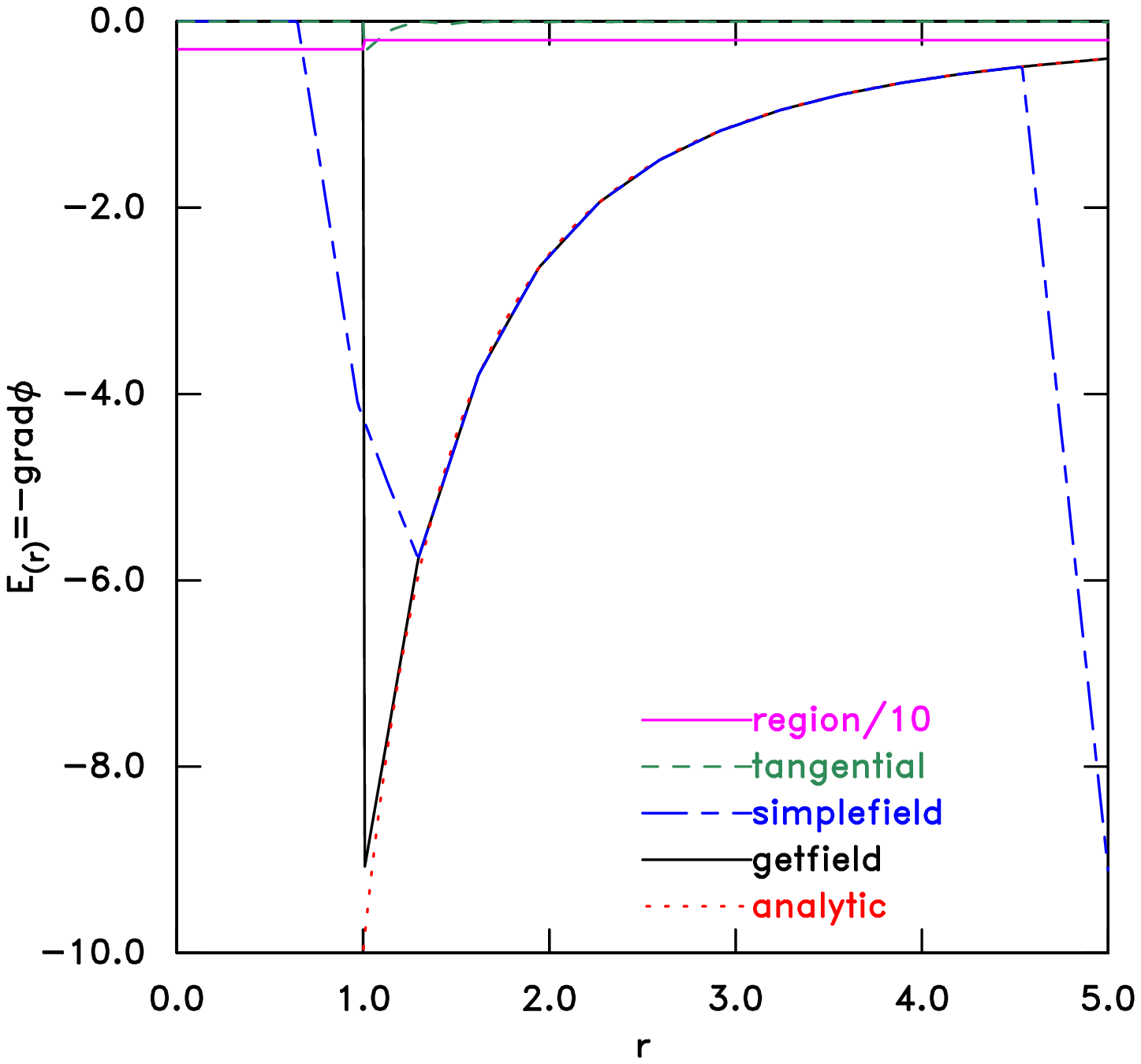}
  \caption{Comparison of different interpolation schemes for $32^3$ mesh.\label{phifield}}
\end{figure}

The values at the mesh nodes tell only the accuracy of the solution
representation at the nodes. Of more interest is the overall accuracy
of the solution when interpolated to positions \emph{away} from the
nodes. Fig.\ \ref{phifield} illustrates this by performing the
interpolations previously described along a line of constant direction
in the $z=0$ plane whose angle is $0.1$ radians ($5.7^o$) to the $x$-axis.
The lines labelled ``getpotential'' and ``getfield'' are the quadratic
interpolations. For comparison, the analytic solution is plotted and
two other types of approximation. That called ``nearest'' simply takes
the potential to be the value at the nearest node of the
mesh. The approximation called ``simple'' is arrived at by a simple
linear interpolation of $\phi$ between nodes, and for ``simplefield'' a linear
interpolation of $\nabla \phi$ based on taking differences of the
$\phi$ nodal values. The ``simple'' $\phi$ and $\nabla \phi$ estimates
therefore ignore the extra information embedded in the boundary
conditions. We notice that although the $\phi$ interpolation in the
bulk region is very good for all approaches, near the boundaries there
are big errors for the simple and nearest approximations. For the
getpotential interpolation, the errors are an order of magnitude
smaller, having a maximum error of $0.145$ along this line, occurring at a
position just outside the inner sphere. This value is typical. (The
``region/10 line'' simply indicates the different regions of the solution.)

The field ($\nabla\phi$) interpolation shows even larger errors in the
simple estimate, because linearly interpolating it from differences in
the vicinity of its discontinuity at $r=1$ produces a piecewise linear
continuous behavior that is far from the analytic form. The quadratic
interpolation \emph{using the boundary information}, by contrast,
accommodates the discontinuity in field and obtains very respectable
approximations even in the vicinity of the boundary. The maximum
field error varies substantially with the angle of the path. In this
case it is about $0.8$ in radial field component and less in
tangential field.

\begin{figure}[ht]
  \includegraphics[height=7cm]{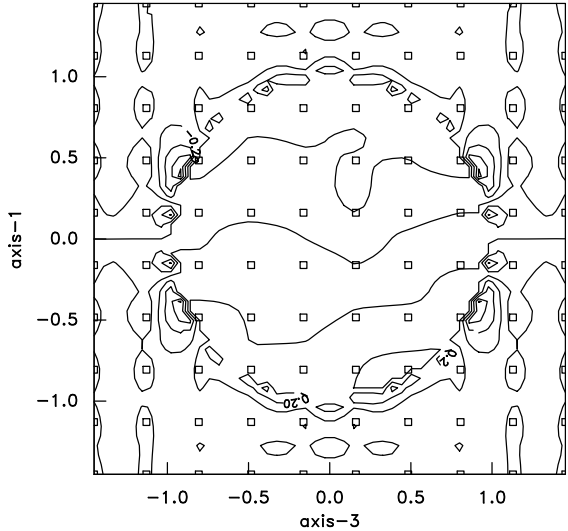}
  \caption{Error in field component in direction-1 within a
    plane of constant $y$ (axis-2). Contour spacing 0.2. (The
    maximum of $|\nabla\phi|$ is 10; so contours are 2\% relative error.)\label{errorcont}}
  \end{figure}

Figure \ref{errorcont} documents the spatial dependence of the error
in the $x$-direction field in a plane $y=$constant ($y\approx0$). The contours of
error are spaced by 0.2.
The positions of the nodes are marked by squares. What we see is that
the field error is less than $0.2$ over most of the domain but locally
near the field discontinuity at $r=1$ larger errors occur that depend
upon position round the boundary.

The scale length of the field and potential variation
immediately outside $r=1$ is 1; the mesh spacing is about
0.3. It is therefore evident that we are obtaining highly favorable
accuracy, typically 10\% maximum local error and 1\% standard deviation, in
field even with rather coarse mesh resolution ($0.3$) of the
scale-length. Moreover using the boundary information we completely
avoid pollution of the field in one region by values in another where
it is different by a discontinuity.

\begin{table}[htp]
\begin{center}
  \begin{tabular}{|l|c|c|c|c|c|}
\hline
Nodes 	        &  $16^3$ &    $32^3$  	& $48^3$ &   $64^3$   &	$128^3$\\
Spacing $dx/r_s$    &  0.63  &    0.31  & 0.21   &    0.156 & 0.078\\\hline
$\Delta\phi_{max}/\phi_{max}$  &  0.020 & 0.0031 & 0.0014 & 0.0008 & 0.0002 \\
$\Delta E_{max}/E_{max}$  & 0.167 &    0.091 & 0.068 & 0.050  &	0.022\\
$SD(\Delta E)/E_{max}$	&  0.020 &    0.0089	& 0.0060 &  0.0022 &	0.0007\\\hline
  \end{tabular}
\end{center}
  \caption{Nodal-potential and field-interpolation relative error scaling with mesh size.\label{errorscale}}  
\end{table}
Table \ref{errorscale} show the maximum nodal (i.e.\ uninterpolated)
potential error $\Delta\phi_{max}/\phi_{max}$, and the maximum local
(interpolated) field error $\Delta E_{max}/E_{max}$ and standard
deviation of the error $SD(\Delta E)/E_{max}$ evaluated over a central cube
of side length 3. These errors decrease with the mesh spacing, as the
number of nodes is increased. The nodal potential error decreases
quadratically with mesh spacing as expected for this second order
scheme. The maximum interpolated field error decreases only linearly,
but the error becomes more and more localized so that the standard
deviation decreases approximately quadratically. By comparison, the
unsatisfactory ``simple'' interpolation (not shown in the table) gives
local field errors whose maximum magnitude is independent of mesh
spacing. The ``getfield'' interpolation
asymptotic \emph{scalings} are the same as a ``nearest'' value interpolation
(using the nearest value of the gradient that is based on values only
from nodes on the same side of any boundary as the point being
interpolated). This is presumably because there are situations near
boundaries where the multilinear interpolation provides no
tangential-field derivative-information, omitting nodes that lie
across a boundary. The \emph{magnitude} of getfield's error, however, is
smaller by a substantial factor (perhaps 3 or 4) than a nearest value
approximation, and the volume of its region of bad approximation is
smaller.

In summary, using a compact difference stencil, Poisson solution and
interpolation of the electric field is implemented with excellent
accuracy and convergence, suitable for use in particle in cell codes
with purely cartesian mesh but essentially arbitrary oblique
boundaries.

\bibliography{numericschemes}

\end{document}